\documentclass[journal]{IEEEtran}

\usepackage{cite}
	
\usepackage[cmex10]{amsmath}
\usepackage{amssymb,amsfonts}
\usepackage[hyphens]{url}
\usepackage{algorithmic,algorithm}
\usepackage{amsthm}

\usepackage{graphicx}
\usepackage[caption=false,font=footnotesize]{subfig}
\usepackage{multirow}
\usepackage{etoolbox}
\usepackage{comment}
\usepackage{bm}
\usepackage{placeins}

\newtoggle{isreport}
\toggletrue{isreport}

\newtoggle{isarxiv}
\togglefalse{isarxiv}

\newtheorem{lemma}{Lemma}

\newtheorem{cor}[lemma]{Corollary}
\newtheorem{thm}[lemma]{Theorem}
\theoremstyle{theorem}
\theoremstyle{theorem}\newtheorem{defn}[lemma]{Definition}

\newcommand{\set}[1]{\left\{#1\right\}}
\newcommand{\abs}[1]{\left|#1\right|}

\newcommand{\normm}[1]{{\left\vert\kern-0.25ex\left\vert\kern-0.25ex\left\vert #1
    \right\vert\kern-0.25ex\right\vert\kern-0.25ex\right\vert}}

\newcommand{\R}{\mathbb R}

\newcommand{\ol}[1]{\overline{#1}}

\newcommand{\calC}{\mathcal{C}}

\newcommand{\calE}{\mathcal{E}}

\newcommand{\calG}{\mathcal{G}}

\newcommand{\calN}{\mathcal{N}}

\newcommand{\calT}{\mathcal{T}}

\usepackage{color}
\usepackage{enumerate}

\definecolor{myred}{RGB}{202,0,32}
\definecolor{myorange}{RGB}{244,165,130}
\definecolor{myviolet}{RGB}{194,165,207}
\definecolor{mycyan}{RGB}{146,197,222}
\definecolor{myblue}{RGB}{5,113,176}
\definecolor{mygreen}{RGB}{127,191,123}
\definecolor{mytile}{RGB}{27,120,55}

\newcommand\cliang[1]{\textcolor{myorange}{[Chen: #1]}}

\newcommand\ale[1]{\textcolor{mygreen}{[Ale: #1]}}

\begin{document}
\title{Line Failure Localization of Power Networks
\\ Part II: Cut Set Outages}

\IEEEoverridecommandlockouts

\author{Linqi Guo, Chen~Liang, Alessandro~Zocca, Steven H.~Low,~and~Adam~Wierman
\thanks{This work has been supported by Resnick Fellowship, Linde Institute Research Award, NWO Rubicon grant 680.50.1529, 
	NSF through grants CCF 1637598, ECCS 1619352, ECCS 1931662, CNS 1545096, CNS 1518941, CPS ECCS 1739355, CPS 154471.}
\thanks{LG, CL, SHL, AW are with the Department of Computing and Mathematical Sciences, California Institute of Technology, Pasadena,
CA, 91125, USA. Email: \texttt{\{lguo, cliang2, slow, adamw\}@caltech.edu}. AZ is with the Department of Mathematics of the Vrije Universiteit Amsterdam, 1081HV, The Netherlands. Email: \texttt{a.zocca@vu.nl}.}}

\maketitle

\begin{abstract}
%
Transmission line failure in power systems prop-agate  non-locally,  making  the  control  of  the  resulting  outages extremely  difficult. 
In Part II of this paper, we continue the study of line failure localizability in transmission networks and characterize the impact of cut set outages. We establish a Simple Path Criterion, showing that the propagation pattern due to bridge outages, a special case of cut set failures, are fully determined by the positions in the network of the buses that participate in load balancing. We then extend our results to  general cut set  outages. In contrast to non-cut outages discussed in Part I whose subsequent line failures are contained within the original blocks, cut set outages typically impact the whole network, affecting the power flows on all remaining lines. We corroborate our analytical results in both parts using the IEEE 118-bus test system, in which the failure propagation patterns exhibit a clear block-diagonal structure predicted by our theory, even when using full AC power flow equations. 
\end{abstract}


\section{Introduction}\label{section:intro}
In Part I of this paper \cite{part1} we establish a spectral representation of power redistribution that precisely captures the Kirchhoff's Laws in terms of the distribution of different families of subtrees in the transmission network. This new representation enables us to precisely characterize how non-cut line outages propagate. In particular, a non-cut outage in a block will only impact the branch power flows on the transmission lines within that block, regardless of whether the outage involves a single line or multiple lines simultaneously. Moreover, a non-cut outage will almost surely impact flows on every remaining line within its block.

\begin{figure}[t]
\centering
\iftoggle{isarxiv}{
\includegraphics[width=.3\textwidth]{arxiv_figs/matrix_block.png}
}{
\includegraphics[width=.3\textwidth]{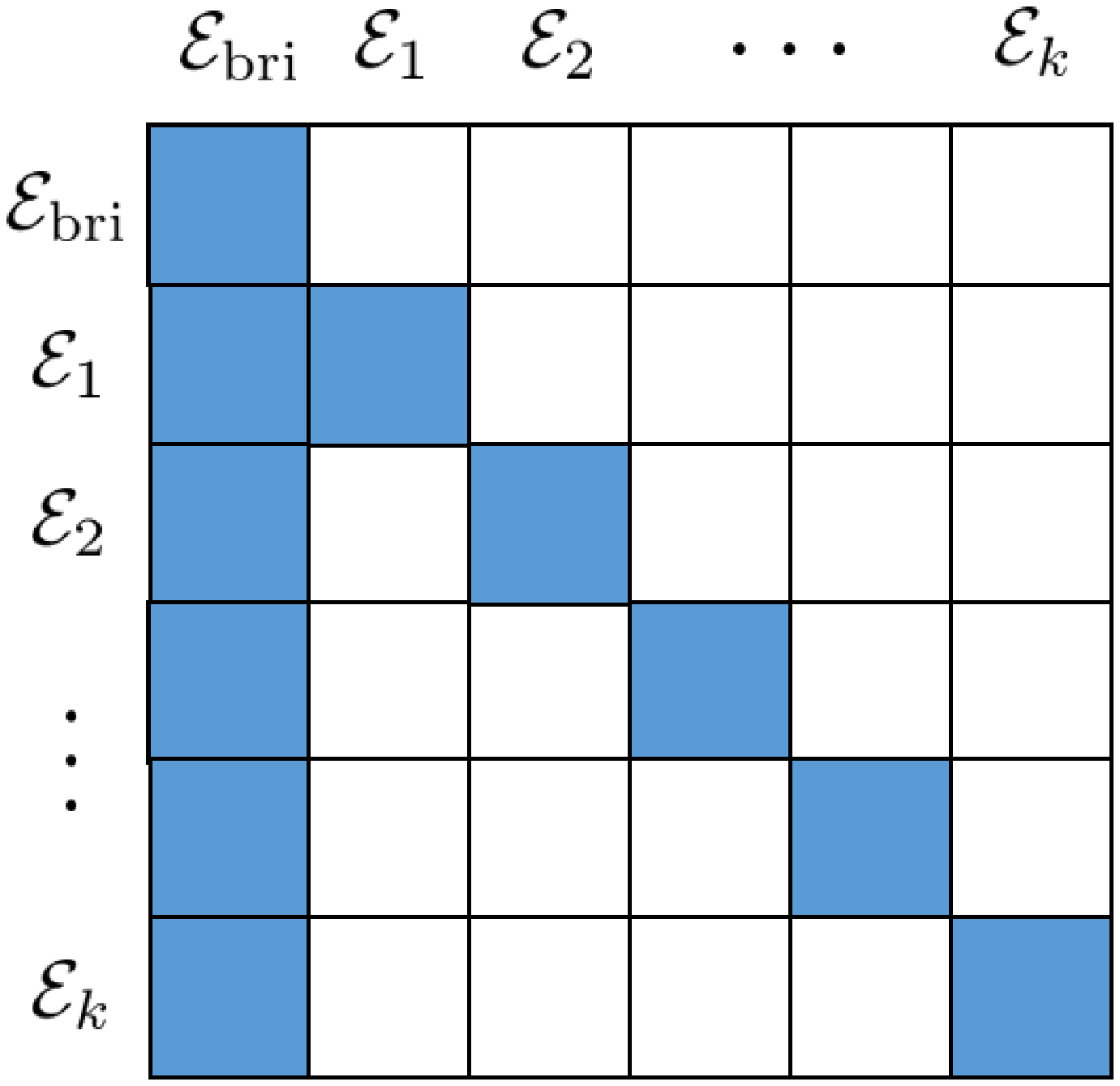}
}
\caption{Non-zero entries of the $K_{l\hat{l}}$ matrix (depicted as the dark blocks) for a graph with non-bridge blocks $\set{\calE_1,\calE_2,\cdots,\calE_b}$ and bridge set $\calE_\text{bri}$.}\label{fig:matrix_block}
\end{figure}

\textbf{Contributions of Part II of this paper:} \emph{We study cut set outages and analytically characterize how such failures impact the remaining lines.} 
Our results demonstrate how the impact of cut set outages propagate globally in a way that depends on both the design of power balancing rules and the network topological structure. This characterization, together with our results from Part I, can be visualized in Fig.~\ref{fig:matrix_block}, where it becomes clear how the block decomposition of a network is linked to the sparsity of the LODF matrix $K_{l\hat{l}}$. This new theory builds on recent work on line outage distribution factors, e.g., \cite{soltan2015analysis, guo2017monotonicity}, and shows that the block decomposition yields an extremely useful representation of these factors.

The formal characterization of single bridge outage is given by a Simple Path Criterion as Theorem \ref{ch:cf; subsec:bridgeoutage; theorem:simplepath} in Section \ref{section:bridge}, which shows that the relative positions of the buses participating in load balancing fully determines how such failures propagate. This result applies to the scenario in which the post-contingency network is disconnected into two or more connected components, known as \textit{islands}, and the original power injections need to be balanced in each island. We then formulate the concept of \emph{participating block} and show that bridge outages typically propagate globally across the network and impact the branch flows on all transmission lines. In Section \ref{section:multi_line}, we extend Theorem \ref{ch:cf; subsec:bridgeoutage; theorem:simplepath} to the case of a general cut set outage, and show that the aggregate impact of such failures can be decomposed into two terms: (a) a first term that captures the effect of power redistribution, which can be decomposed in accordance to the blocks where the failures occur and is fully characterized by our study in Part I; and (b) a second term that describes the impact of the power balancing rule and generalizes the case of a single bridge failure, capturing how the system handles disconnected components.


Results in Parts I and II of this paper provide a complete characterization of line failure propagation and is illustrated in Section \ref{section:case_study} using IEEE 118-bus test network. In particular, we show that the LODF matrix has a clear block-diagonal structure predicted by our theory, even when using full AC power flow equations.

\section{Islanding Model}\label{section:IslandModel}
In contrast to Part I where we focus on simultaneous line outages that do not disconnect the network, we consider now the case in which the set of initial line outages disconnects the network into two or more connected sub-networks, called \emph{islands}. We refer to such contingency as a \emph{cut set outage} and, in the special case in which the cut set consists of a single line, as a \emph{bridge outage}.
We remark that, in practice, such islands can be created accidentally by line outages, but also deliberately as a defensive action to prevent a disturbance/attack from propagating 
across the entire network infrastructure \cite{Li2005}.

\subsection{Islands and Cut Set Outages}
In this section we present a detailed model for a single island that is necessary for our analysis of cut set outages in the later sections. More specifically, we fully characterize the impact of a bridge outage in Section \ref{section:bridge} and of a general cut set outage in Section \ref{section:multi_line}.

We adopt the same notations as in Part I, which we now summarize. Let $\calG'=(\calN', \calE')$ denote the pre-contingency network and consider a subset of lines $F'\subset \calE'$ that is a cut set of $\calG'$ and denote by $\calG_1, \dots, \calG_k$ the multiple islands created by the removal of lines in $F'$. Let us focus on one of these islands, say $\calG=(\calN, \calE)$, where $\calN \subseteq \calN'$ is the set of buses that belong to the island and $\calE \subseteq \calE'$ is the set of lines that, \emph{pre-contingency}, have both endpoints inside the island. From the viewpoint of the island $\calG$, the lines in the cut set $F'$ can then be classified into three types, $F'=F_{\text{external}} \cup F_{\text{tie}} \cup F$, as follows: 
\begin{itemize}
	\item $F_{\text{external}} := \set{\hat l = (\hat i,\hat j) \in F': \hat i \notin \calN, \hat j \notin \calN}$ is the set of tripped \emph{external lines} with neither endpoints in the island $\calG$;
	\item $F_{\text{tie}}:= \set{\hat l = (\hat i,\hat j) \in F': \hat i \notin \calN, \hat j \in \calN}$ is the set of tripped \emph{tie lines} with exactly one endpoint (denoted by as $j(\hat l)$) in the island $\calG$;
	\item $F:= \set{\hat l = (\hat i,\hat j) \in F': \hat i \in \calN, \hat j \in \calN}$ is the set of tripped \emph{internal lines} with both endpoints inside the island $\calG$.
\end{itemize}
Note that external line outages do not have a direct impact on the island $\calG$ since the post-contingency operations and power flow equations are decoupled in each island and thus can be ignored. Therefore, without loss of generality, we henceforth assume $F' = F_{\text{tie}} \cup F$. Since the graph $\calG=(\calN, \calE)$ describes the pre-contingency topology of the island, its edge set $\calE$ includes the tripped internal lines $F$, but not the tie lines $F_{\text{tie}}$. We refer to the lines that are still active in the island post-contingency as \textit{surviving lines} and denote their collection as $-F:=\calE \setminus F$. The post-contingency island is thus fully described by the graph $(\calN, \calE \setminus F)$, which is connected by construction. In particular, the tripped internal lines $F$ is a non-cut set outage of island $\calG$. 

Designate any bus in $\calN$ to be the reference bus for $\calG$ and, without loss of generality, assume it is bus $n$. Let $B$ denote the susceptance matrix associated with the island $\calG$, $C$ its incidence matrix, $L:=CBC^T$ its Laplacian matrix, and define $A$ in terms of the reduced Laplacian matrix $\ol{L}$ as in Part I of this paper:
\begin{align*}
A = \begin{bmatrix}
\ol{L}^{-1} & \bm{0} \\
\bm{0} & 0
\end{bmatrix}.
\end{align*}
Let $f:=(f_{\hat l}, \hat l\in\calE)$ and $f^{\text{tie}}:=(f_{\hat l},\hat l \in F_{\text{tie}})$ be the pre-contingency branch flows on the lines inside the island $\calG$ and on the tie lines respectively. We adopt the convention that $f_{\hat l} > 0 $ for a line $\hat l=(\hat i, \hat j)$ if power flows from bus $\hat i$ to bus $\hat j$ over line $\hat l$. In particular, for a tie line $\hat l \in F_{\text{tie}}$, $f_{\hat l} > 0$ implies that pre-contingency the island imports power over line $\hat l$ and $f_{\hat l}<0$ if it exports power over $\hat l$.

If the pre-contingency branch flow $f_{\hat l} = 0$ on any tie line or internal line $\hat l \in F'$, then its tripping
has no impact on the post-contingency branch flows in this island, as modeled by the DC power flow equations.
We therefore assume without loss of generality that the pre-contingency branch flows $f_{\hat l}\neq 0$ for all 
tripped lines $\hat l\in F'$ (otherwise, remove $\hat l$ from $F'$ and the surviving island $\calG$). 

\subsection{Pre-contingency Injections and Branch Flows}
Let $p := (p_k, k\in  \calN)$ denote the pre-contingency injections in the buses of island $\calG$. The effect of pre-contingency tie line branch flows on the island $\calG$ can be modeled by additional injections $f_{\hat l}$ at buses $j(\hat l)$:
\begin{eqnarray}
    \Delta p_\text{tie} :=  \sum_{\hat l\in F_\text{tie}} f_{\hat l} \ e_{j(\hat l)},
\label{ch:cf; subsec:bridgeoutage; eq:P}
\end{eqnarray}
where $e_{j(\hat l)}$ is the standard unit vector of size $n:=|\calN|$. Hence, for the purpose of computing pre-contingency branch flows in island $\calG$, the injections can be taken to be $p + \Delta p_{\text{tie}}$.
Let $f_{F} := (f_l, l\in F)$ denote the pre-contingency branch flows on lines in $F$,  $f_{-F} := (f_l, l\in \calE\!\setminus\! F)$ those on the surviving lines in $\calG$, and $f := (f_{-F}, f_{F})$. Partition the matrices $(B, C)$ according to the two sets of lines, $F$ and $-F=\calE \!\setminus\! F$, as follows:
\begin{eqnarray*}
    B =: \begin{bmatrix}
    	B_{-F} & 0 \\
    	0 & B_{F}
    \end{bmatrix}, \qquad
    C =: \begin{bmatrix} C_{-F}, & C_F \end{bmatrix}.
\end{eqnarray*}
From \eqref{ch:cf; subsec:bridgeoutage; eq:P} it follows that $(f, \theta)$ satisfies the DC power flow equations on the
pre-contingency island $\calG = (\calN, \calE)$:
\begin{subequations}
\begin{IEEEeqnarray}{rCl}
    p +  \Delta p_\text{tie} & = & Cf  = C_{-F} f_{-F} + C_{F} f_{F},\\
    f & = &  BC^T \theta,
\label{ch:cf; subsec:bridgeoutage; eq:DCPFpre.2}
\end{IEEEeqnarray}
where $\theta$ are the pre-contingency voltage angles. 
\label{eqn:DCPF_pre}
\end{subequations}

\subsection{Post-contingency Injections and Branch Flows}
The effect of tie line outages $F_{\text{tie}}$ on  island $\calG$ can be modeled as the loss of the injections
$\Delta p_\text{tie}$ at the endpoints of the tie lines. The pre-contingency injections are then unbalanced over the island $\calG$ with a total imbalance equal to
\begin{eqnarray*}
    \sum_{k\in\calN} p_k = -\sum_{\hat l\in F_\text{tie}} f_{\hat l}. 
\end{eqnarray*}
Post contingency, there is a surplus if the island net exports power pre contingency and a shortage otherwise, depending on the sign of this total imbalance. A balancing rule $\mathbb G$ is invoked to rebalance power in the island by adjusting
injections (generators and/or loads) in response to the contingency. A popular balancing rule, which we name \emph{proportional control}, prescribes how to share the imbalance proportionally among a set of participating buses. More specifically. a proportional control $\mathbb G_\alpha$ is defined by a nonnegative vector $\alpha := (\alpha_k, k\in \calN)$ such that $\sum_{k \in  \calN}\, \alpha_k = 1$, with the interpretation that, post contingency, each bus $k\in  \calN$ adjusts its injection by the amount:
\begin{align}
    \mathbb G_\alpha: \quad \Delta p_k := -\alpha_k \sum_{k\in\calN} p_k =  \alpha_k \sum_{\hat l\in F_\text{tie}} f_{\hat l}, \quad k \in \calN.
    \label{ch:cf; subsec:bridgeoutage; eq:Galpha.1a}
\end{align}
We call a bus $k$ \emph{participating} if $\alpha_k>0$. By design, all participating buses adjust their injections in the same direction. Examples of proportional control $\mathbb G_\alpha$ include participation factors used in automatic generation control or economic dispatch \cite[Chapter 3.8]{WoodWollenbergSheble2014}, or equal sharing of total imbalance among participating buses~\cite{bernstein2014power,soltan2015analysis}; see also \cite{bienstock2010n-k,bienstock2011control,bernstein2014vulnerability,zocca2020optimization}. In this paper we focus solely on proportional control $\mathbb G_\alpha$. For a different class of balancing rules in which the post-contingency injections are determined as a solution of an optimization problem  to minimize the number of buses involved, see \cite{LiangGuo2020}.

Under $\mathbb G_\alpha$, the vector of injection adjustments are then
\begin{eqnarray}
    \Delta p_\alpha := ( \Delta p_k, k\in \calN ) :=  \sum_{\hat l\in F_\text{tie}} f_{\hat l} \ \sum_{k \in \calN}\, \alpha_k e_k,
    \label{ch:cf; subsec:bridgeoutage; eq:Galpha.1b}
\end{eqnarray}
where $e_k$ is the standard unit vector in $\R^n$. The post-contingency injections are thus $p + \Delta p_\alpha$.
Since $\sum_{k \in  \calN}\, \alpha_k = 1$, the identity $\sum_{k\in\calN} ( p_k + \Delta p_k ) = 0$ holds for any initial injection vector $p$, which means that post-contingency power injections are always rebalanced under the proportional control $\mathbb G_\alpha$.


Let $(\tilde f_{-F}, \tilde \theta)$ denote the post-contingency branch flows and voltage angles, which satisfy the DC power flow equations on the post-contingency network $(\calN, \calE\!\setminus\! F)$:
\begin{subequations}
\begin{IEEEeqnarray}{rCl}
    p + \Delta p_\alpha  & = & C_{-F} \tilde f_{-F},\\
    \tilde f_{-F}  & = & B_{-F} C_{-F}^T \tilde \theta.
\label{ch:cf; subsec:bridgeoutage; eq:DCPFpost.2}
\end{IEEEeqnarray}
For post-contingency network, we denote the Laplacian matrix by $L_{-F}:=C_{-F}B_{-F}C_{-F}^T$ and define the matrix $A_{-F}$ correspondingly in terms of the reduced Laplacian matrix $\ol{L}_{-F}$.
\end{subequations}

In the next two sections we use this island model to analyze line outage localization within the island $\calG$ under the proportional control $\mathbb G_\alpha$.

\section{Bridge Outage}\label{section:bridge}

In this section we focus on the case of a single bridge $\hat l$ outage, i.e., $F_{\text{tie}} := \{\hat l\}$, and no internal line outages, i.e., $F := \emptyset$. In the next section we will then extend the results to a cut set outage where the post-contingency branch flows in island $\calG$ are impacted by both internal line outages in $F\subset \calE$
and by tie line outages in $F_\text{tie}$.  Since $F$ is not a cut set in $\calG$ we study this impact by combining the analysis of a non-cut outage in Part I of the paper and that of a bridge outage.

Consider a single bridge $\hat l$ outage with pre-contingency branch flow $f_{\hat l}$ that disconnects the network into two islands. Focus on one of them, say $\calG$. Let $\hat j:=j(\hat l)$ be the endpoint of $\hat l$ in island $\calG$. In this case, 
\begin{align}
    \Delta p_\text{tie} = f_{\hat l} \, e_{\hat j}.
\label{ch:cf; subsec:bridgeoutage; eq:P-l}
\end{align}

Post contingency, the injections are changed from $p+\Delta p_{\text{tie}}$ to $p+\Delta p_{\alpha}$ under the proportional control $\mathbb{G}_\alpha$ as defined in \eqref{ch:cf; subsec:bridgeoutage; eq:Galpha.1b}. Note that since $F=\emptyset$, we have $-F=\calE$, $C_{-F}=C$, $B_{-F}=B$ and $A_{-F}=A$. Therefore, the post-contingency branch flows $\tilde{f}$ in the island $\calG$ are given by
\begin{IEEEeqnarray}{rCl}
    \tilde f & = & BC^T A \left( p + \Delta p_\alpha \right) \nonumber \\
    & = & BC^T A \left( p + \Delta p_{\text{tie}} + \Delta p_\alpha - \Delta p_{\text{tie}} \right)  \nonumber \\
    & = &  f + BC^TA  \left( \Delta p_\alpha - \Delta p_\text{tie} \right).
\label{ch:cf; subsec:bridgeoutage; eq:tildeP}
\end{IEEEeqnarray}
Using \eqref{ch:cf; subsec:bridgeoutage; eq:Galpha.1b}, \eqref{ch:cf; subsec:bridgeoutage; eq:P-l}, and the fact that $\sum_{k\in\calN} \alpha_k = 1$, we get
\begin{IEEEeqnarray}{rCl}
    \Delta p_\alpha - \Delta p_\text{tie} & = &  
    f_{\hat l}\, \sum_{k\in\calN} \alpha_k e_k - f_{\hat l}\, e_{\hat j} \sum_{k\in\calN} \alpha_k \nonumber \\
    & = & f_{\hat l}\, \sum_{k\in \calN} \alpha_k  \left( e_k - e_{\hat j} \right).
\label{ch:cf; subsec:bridgeoutage; eq:Deltap.1}
\end{IEEEeqnarray}
Substituting this expression in \eqref{ch:cf; subsec:bridgeoutage; eq:tildeP} gives 
\begin{IEEEeqnarray*}{rCl}
	\tilde f - f & = & f_{\hat l} \,  \sum_{k\in \calN} \alpha_k \,  BC^T A  \left( e_k - e_{\hat j} \right).
\end{IEEEeqnarray*}
Recalling that $f_{\hat l}\neq 0$ by assumption, for any line $l = (i,j) \in \calE$ we obtain
\begin{IEEEeqnarray*}{rCl}
    \frac{\tilde f_l - f_l } {f_{\hat l}} & = & 
    \sum_{k\in \calN} \alpha_k \, B_l \left( A_{ik} + A_{j\hat j} - A_{i\hat j} - A_{jk} \right) \\
    & = & \sum_{k: \alpha_k>0} \alpha_k \, D_{l, k\hat j},
\end{IEEEeqnarray*}
where $D_{l, k \hat j}$ is the power transfer distribution factor (PTDF) for island $\calG$ discussed in Part I of the paper.
Therefore, the branch flow change on line $l$ is the superposition of impacts due to injecting $\alpha_k f_{\hat l}$ at participating buses $k\in \calN$ and withdrawing them at bus $\hat j$.
We can thus extend the definition of line outage distribution factor (LODF) $K_{l\hat l}$ to allow bridge outages as follows: given a bridge outage $\hat l$, under the proportional control $\mathbb G_\alpha$ for all $l\in \calE$ we have 
\begin{eqnarray}
    K_{l\hat l} := \frac{ \tilde f_l - f_l } { f_{\hat l} }  =  \sum_{k: \alpha_k>0}  \alpha_k\, D_{l, k\hat j}.
\label{ch:cf; subsec:bridgeoutage; eq:Klhatl.1}
\end{eqnarray}
Remember that $\hat l\not\in \calE$ in the island model that we consider here.

The next result is analogous to the Simple Cycle Criterion (see Theorem 8 in Part I of this paper) for non-bridge outages. It states that all ($\mu$-almost surely), and only, lines on a simple path\footnote{A simple path is a path that visits each node at most once.} between bus $\hat j$ and a participating bus $k\in\calN$ with $\alpha_k>0$ will be impacted by the bridge $\hat l$ outage.
\begin{thm}[Simple Path Criterion: bridge outage]
\label{ch:cf; subsec:bridgeoutage; theorem:simplepath}
For a single bridge $\hat l$ outage, under the proportional control $\mathbb G_\alpha$, for every line $l$ in the island $\calG$, $K_{l\hat l} \neq 0$ ``if'' and only if there exists a simple path in $\calG$ 
that contains $l$ from $\hat j$ to a participating bus $k\in \calN$.
\end{thm}
See Appendix \ref{proof:balancing_bus} for a proof. Note that if $\hat j$ (the endpoint of $\hat l$ in $\calG$) is the only participating bus, then $K_{l \hat l}=0$ for all $l\in \calE$. 

Denote by $\calE_1$, $\dots$, $\calE_b$ the unique block decomposition of $\calG$.\footnote{See Part I of the paper for more details on block decomposition.} We say a block $\calE_k$ is \emph{on a simple path between bus $\hat j$ and a participating bus $i \in \calN$} if there is a simple path between bus $\hat j$ and bus $i$ with $\alpha_i>0$ that contains a line $l\in \calE_k$. From Theorem \ref{ch:cf; subsec:bridgeoutage; theorem:simplepath} we can deduce the following localization property after a single bridge $\hat l$ outage in terms of the block structure of the island $\calG$. 
\begin{cor}[Bridge outage]
\label{ch:cf; subsec:bridgeoutage; corollary:bridgeoutage}
Under the proportional control $\mathbb G_\alpha$ \eqref{ch:cf; subsec:bridgeoutage; eq:Galpha.1b}, for any block $\calE_k$ of the island $\calG$ the following statement hold:
\begin{enumerate}
\item $\tilde f_l = f_l$ for all lines $l$ in block $\calE_k$ if $\calE_k$ is not on a simple path between bus $\hat j$ and a participating bus.
\item Conversely, $\tilde f_l \neq f_l$ for all lines $l$ in $\calE_k$ ``if'' $\calE_k$ is on a simple path between bus $\hat j$ and a participating bus.
\end{enumerate}
\end{cor}
%
Corollary \ref{ch:cf; subsec:bridgeoutage; corollary:bridgeoutage} shows that the positions of participating buses play an important role in distributing the power imbalance across the network. In particular, the proportional control $\mathbb{G}_\alpha$ almost surely changes the branch flow on every line that lies in a path from the failure endpoint $\hat j$ to the set of participating buses. As a result, if $l$ is a bridge connecting two sub-networks  $\calG_1$ and $\calG_2$ post contingency (each of which contains one or more blocks), assuming $\hat j\in\calG_1$, then $\Delta f_{l}\neq 0$ ``if'' and only if there is a participating bus in $\calG_2$ since a path from $\hat j$ to any node in $\calG_2$ must pass through the bridge $l$. If $l$ is not a bridge, i.e. $l$ belong to a non-bridge block,  then we can devise a simple sufficient condition for $\Delta f_{l}\neq 0$ using \emph{participating blocks}, defined as follows:
\begin{defn}\label{defn:participating}
Consider an island $\calG$ with block decomposition $\calE_1, \calE_2, \dots, \calE_b$ operating under proportional control $\mathbb{G}_\alpha$ with a set $\calN_\alpha:=\set{i \in \calN: \alpha_i > 0}$ of participating buses. A non-bridge block is said to be a \emph{\textbf{participating block}} if there is a non-cut vertex in this block that is a participating bus. \end{defn}
If all generators participate in AGC and load-side participation is implemented at all load buses, then every node in the network is a participating bus and, hence, every block is participating.

The following result, whose proof is presented in Appendix \ref{proof:participating_region}, shows that if a non-bridge block is participating, then all lines inside it are impacted when the original bridge $\hat l$ is disconnected.
\begin{cor}\label{lemma:participating_region}
Consider a bridge outage $\hat l$ with non-zero branch flow $f_{\hat l} \neq 0$ and let the block decomposition of island $\calG$ be $\calE_1, \dots, \calE_b$. If $\calE_k$ is a participating block, then $\Delta f_{l}\neq 0$ $\mu$-almost surely for any $l \in \calE_k$, i.e., $\mu(\Delta f_{l}\neq 0) = 1.$
\end{cor}

\section{Cut Set Outage}\label{section:multi_line}

We now extend our results to a cut set outage $F'$. Consider an island $\calG=(\calN, \calE)$ and, as before, partition the tripped lines into tie lines and internal lines, i.e. $F'=F\cup F_{\text{tie}}$. 

The impact on post-contingency branch flows is a superposition of the impact of internal line outages in $F$, weighted by generalized line outage distribution factor (GLODF) with multiple tripped lines $K^F$, and the impact of tie line outages in $F_\text{tie}$, weighted by the proportional control $\alpha_k$ as well as PTDF of the post-contingency network, as stated in the following theorem.

\begin{thm}
\label{ch:cf; sec:localization; subsec:cutset; thm:cutset}
Given an island $\calG=(\calN,\calE)$ with a cut set outage $F'$, under the proportional control $\mathbb G_\alpha$ \eqref{ch:cf; subsec:bridgeoutage; eq:Galpha.1b} the branch flow changes on the surviving lines in $-F=\calE \!\setminus\! F$ are given by
\begin{equation}
    \Delta f_{-F} = K^F f_F + \sum_{\hat l \in F_{\textrm{tie}}} f_{\hat l} \sum_{k\in\calN} \alpha_k \hat D^F \left( e_k - e_{j(\hat l)} \right),
\label{ch:cf; sec:localization; subsec:cutset; eq:cutset.1}
\end{equation}
\begin{subequations}
where $K^F:= B_{-F}C_{-F}^TAC_{F}\left( I - B_FC_F^T A C_F \right)^{-1}$ is the GLODF\footnote{See Part I of the paper for more details on  GLODFs.} of island $\calG$ with a non-cut outage $F$, and $\hat D^F$ is the PTDF for the post-contingency network $(\calN, \calE \setminus F)$ defined as
\begin{equation}
    \hat D^F:=B_{-F}C_{-F}A_{-F},
\label{ch:cf; sec:localization; subsec:cutset; eq:DF.1a}
\end{equation}
which can equivalently be expressed in terms of the pre-contingency island $\calG$ as
\begin{equation}
    \hat D^F = \left(B_{-F}C_{-F}^T + K^F B_F C_F^T \right)A.
    \label{ch:cf; sec:localization; subsec:cutset; eq:DF.1b}
\end{equation}
\end{subequations}
\end{thm}
The theorem reduces to~\eqref{ch:cf; subsec:bridgeoutage; eq:Klhatl.1} for a single bridge $\hat l$ outage, that is when $F=\emptyset$ and $F_{\text{tie}} = \{\hat l\}$. 
When the cut set outage contains both internal line outages in $F$ and tie line outages in $F_{\text{tie}}$, the post-contingency branch flows depend in an intricate way on both types of outages. Our theorem makes this relationship explicit and precise:
\begin{enumerate}[(a)]
\item 
The first term on the right-hand side of \eqref{ch:cf; sec:localization; subsec:cutset; eq:cutset.1} represents the impact of the outage of a non-cut set $F$ of internal lines in $\calG$ through the GLODF $K^{F}$ of island $\calG$. If there are no tie line outages $F_{\text{tie}} = \emptyset$, then the formula reduces to the GLODF for a non-cutset outage as discussed in Part I.
\item The second term on the right-hand side of \eqref{ch:cf; sec:localization; subsec:cutset; eq:cutset.1}
represents the impact of the proportional control $\mathbb G_\alpha$ in response to tie line outages in $F_\text{tie}$.
If there are no internal line outages $F = \emptyset$ then the formula reduces to
\begin{equation*}
\Delta f_{-F} = \sum_{\hat l \in F_{\text{tie}}} f_{\hat l} \sum_{k\in\calN} \alpha_k \hat D^F \left( e_k - e_{j(\hat l)} \right),
\end{equation*}
which generalizes \eqref{ch:cf; subsec:bridgeoutage; eq:Klhatl.1} from a single bridge $\hat l$ outage to a cut set $F_{\text{tie}}$ outage with multiple tripped tie lines.  In this case the impact of simultaneous outage of a set $F_\text{tie}$ of tie lines is simply the sum of the impacts of single-bridge outages \emph{as if} $\hat l$ is a bridge incident on $\calG$. The expressions for $\hat D^F$ \eqref{ch:cf; sec:localization; subsec:cutset; eq:DF.1a} and \eqref{ch:cf; sec:localization; subsec:cutset; eq:DF.1b} in terms of pre- and post-contingency network trivially coincide when $F=\emptyset$.

\item When both $F\neq \emptyset$ and $F_{\text{tie}}\neq \emptyset$, the effect of their outages on post-contingency branch flows can be interpreted in terms of either the pre-contingency network $A$ or the post-contingency network $A_{-F}$, through the expressions for both $K^F$ and $\hat D^F$. 
In particular the expression for $\hat D^F$ in terms of the pre-contingency network $A$ has two components. The first component $B_{-F}C_{-F}^T A$ says that the injection adjustments due to the proportional control $\alpha_k$ will change the branch flows on the surviving lines in $-F$ according to PTDF.
The second component $K^F B_F C_F^T A$ says that the injection adjustments are first mapped to flow changes on lines in $F$ through PTDF $B_FC_F^TA\left( e_k - e_{j(\hat l)} \right)$ in the \emph{pre-contingency} network and they are then mapped to flow changes on surviving lines in $-F$ through $K^F$ when lines in the non-cut set $F$ are disconnected.

\item According to the expression \eqref{ch:cf; sec:localization; subsec:cutset; eq:DF.1a} for $\hat D^F$, the post-contingency network integrates both effects: changes to post-contingency branch flows are the sum of the impact of injection adjustments under proportional control $\alpha_k$ through PTDF $B_{-F}C_{-F}^TA_{-F}\left( e_k - e_{j(\hat l)} \right)$ on the \emph{post-contingency} network.
\end{enumerate}

Considering a surviving line $l\in -F$, the flow change is given by, in terms of pre-contingency island $\calG$:
\begin{IEEEeqnarray}{rCl}
\Delta f_l & = & \sum_{\tilde{l} \in F} K_{l\tilde{l}}^F f_{\tilde{l}}  \nonumber\\
&&  +\sum_{\hat l\in F_{\text{tie}}} f_{\hat l} \sum_{k\in\calN} \alpha_k \Big(D_{l, kj(\hat l)} + \sum_{\tilde{l} \in F} K_{l \tilde{l}}^F D_{\tilde{l}, kj(\hat l)} \Big) \nonumber\\
& = & \sum_{\tilde{l} \in F} K_{l\tilde{ l}}^F \Big(f_{\tilde{l}} + \sum_{\hat l \in F_{\text{tie}}} \sum_{k\in\calN} \alpha_kD_{\tilde{l}, kj(\hat l)} f_{\hat l} \Big) \nonumber\\
& &+\sum_{\hat l\in F_{\text{tie}}} f_{\hat l} \sum_{k\in\calN} \alpha_k D_{l, kj(\hat l)}.
\label{eqn:flow_change_cut}
\end{IEEEeqnarray}
From Theorem 12 in Part I (which covers the case of a non-cut outage) it follows that the first term is zero if $l$ is not in the same block as any disconnected internal line $\hat l\in F$ (since $K_{l\hat l}^F=0$). Applying the Simple Path Criterion in Theorem \ref{ch:cf; subsec:bridgeoutage; theorem:simplepath} implies that the second term is zero if, for every tie line $\hat l\in F_{\text{tie}}$ and every participating bus $k\in \calN$ with $\alpha_k>0$, $l$ does not lie on any simple path in $\calG$ connecting $j(\hat l)$ and $k$. We can thus derive the following localization property in terms of the block decomposition of $\calG$. 
\begin{cor}[Cut set outage]
For a cut set $F'$ outage, under the proportional control $\mathbb G_\alpha$ \eqref{ch:cf; subsec:bridgeoutage; eq:Galpha.1b}, for any surviving line $l\in -F$ in the island $\calG$, $\Delta f_l = 0$ if the unique block $\calE_k$ containing $l$ satisfies both the following conditions:
\begin{itemize}
\item $\calE_k$ contains no disconnected internal line $\hat l\in F$; and 
\item For every tie line $\hat l\in F_{\text{tie}}$ and every participating bus $i\in \calN$ with 
	$\alpha_i>0$, $\calE_k$ is not on a simple path in $\calG$ between $j(\hat l)$ and $k$.
\end{itemize}
\end{cor}
The converse in general is false because, even when both terms in \eqref{eqn:flow_change_cut} are nonzero, they may cancel with each other, resulting in $\Delta f_l = 0$.

\section{Case Studies}\label{section:case_study}
\begin{figure}[t]
\centering
\subfloat[DC\label{fig:LODF-118.1}]{\includegraphics[width=.45\textwidth]{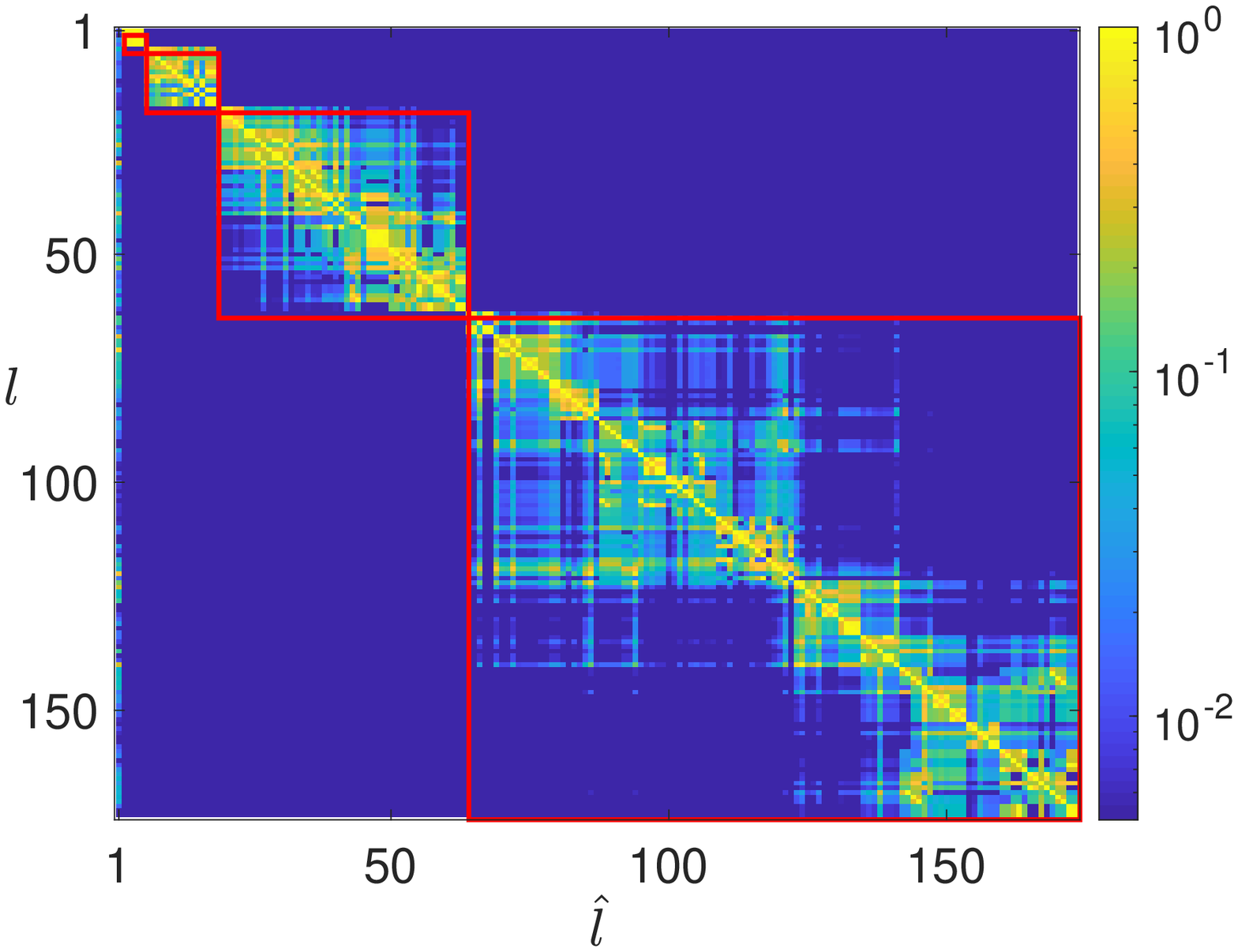}}\\
\subfloat[AC\label{fig:LODF-118.2}]{\includegraphics[width=.45\textwidth]{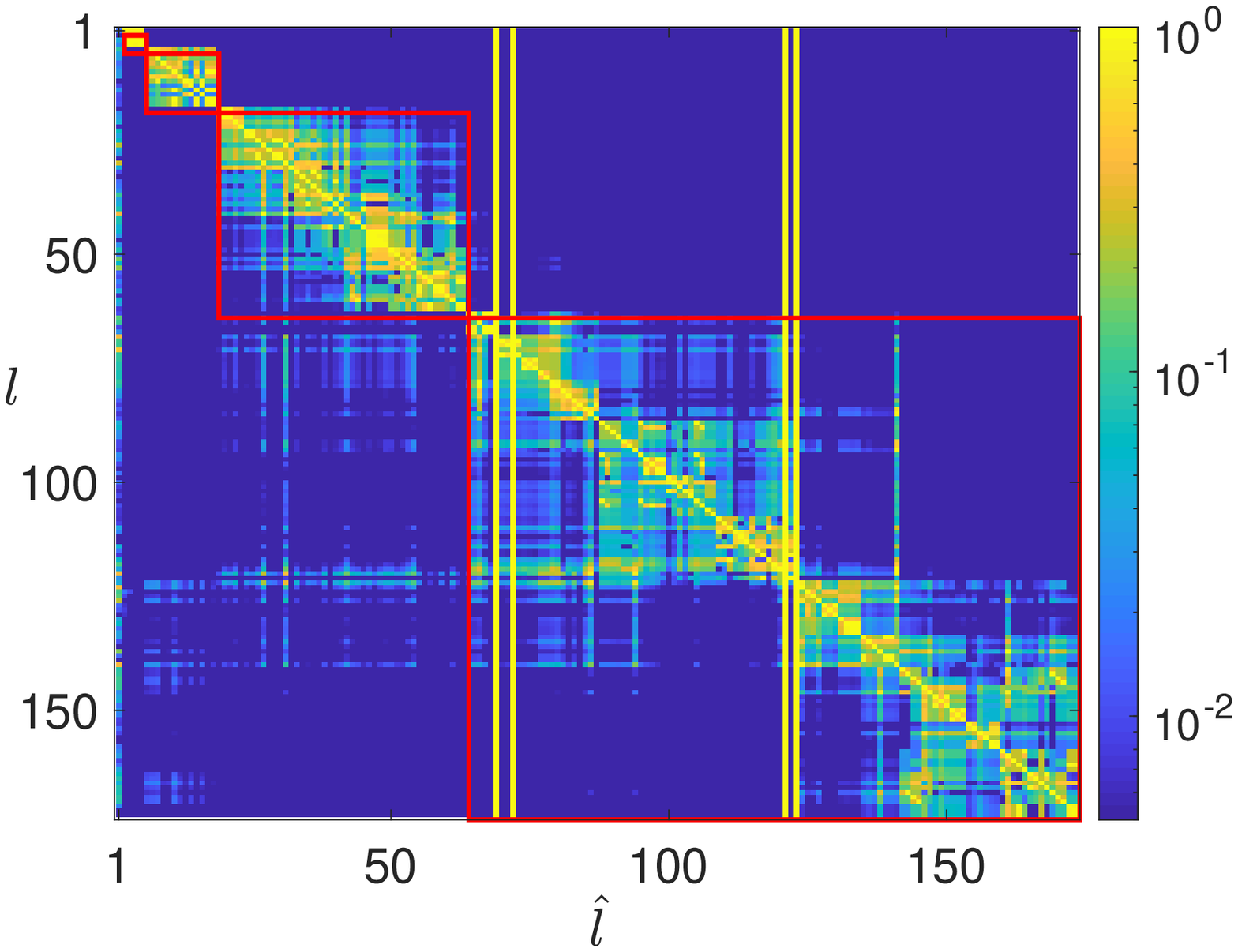}}\\
\caption{The LODF matrix (reporting the absolute values of the distribution factors) of IEEE 118-bus network calculated under (a) DC and (b) AC power flow model. The four yellow solid line in (b) represent four transmission lines whose failure lead to non-convergent AC power flow equations. The red rectangles indicate blocks of the network.}
\label{fig:LODF-118}
\end{figure}
\begin{figure*}[ht]
\centering
\includegraphics[width=.9\textwidth]{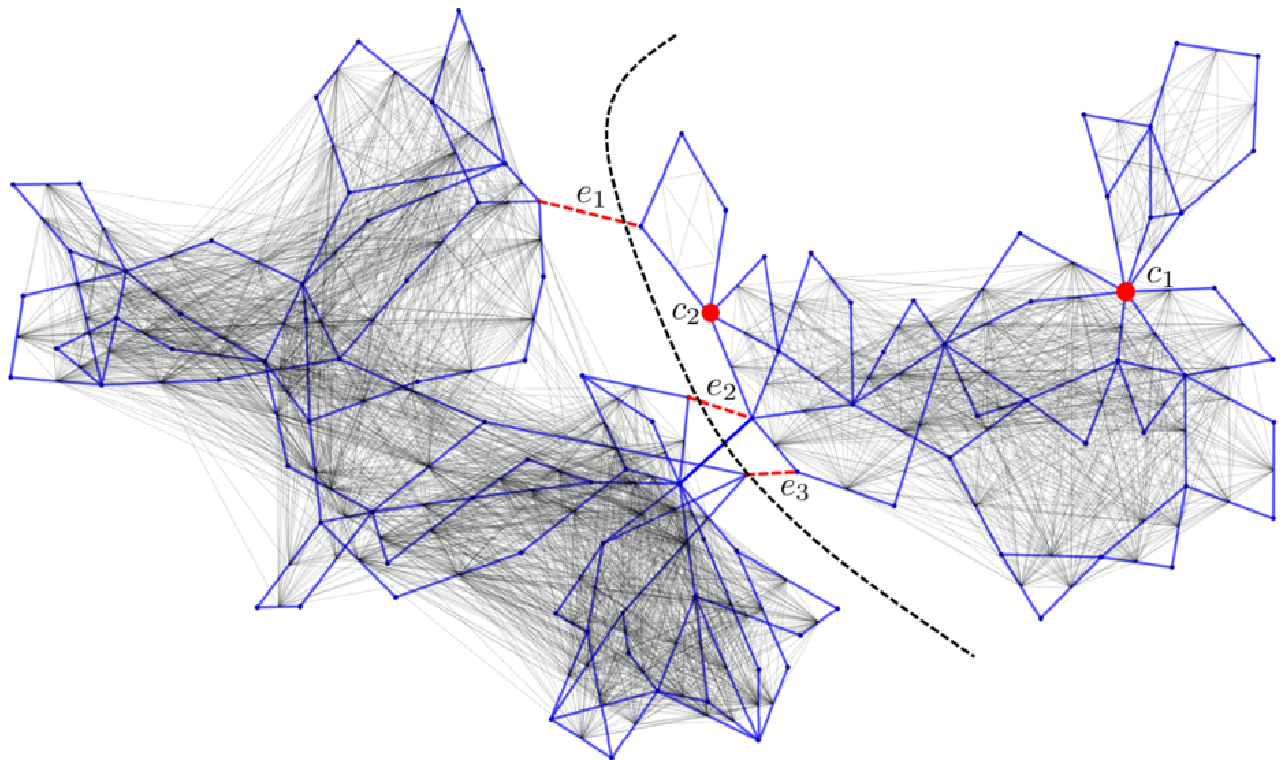}
\caption{Influence graphs on the IEEE 118-bus test network after switching off lines $e_1$, $e_2$ and $e_3$. Blue edges represent physical transmission lines and grey edges represent connections in the influence graph. The black dashed line and the red solid points indicate the failure propagation boundary defined by the blocks.}\label{fig:inf_graph}
\end{figure*}

Theorem 7 in Part I and Theorem~\ref{ch:cf; sec:localization; subsec:cutset; thm:cutset} in Part II summarize the mathematical theory that characterizes the patterns of line failure propagation in power systems. More specifically, the failure localizability depends critically on the block decomposition of a power network. In this section, we demonstrate these localizability properties through simulations using the IEEE 118-bus test network. 

 
\subsection{IEEE 118-bus Network}
This IEEE test network consists of 118 nodes and 186 edges and has a block decomposition with two non-bridge blocks: one giant block with 164 edges, and a smaller one with 13 edges. There are trivial ``dangling" appendages each of which connects a single node to the giant block\footnote{It should be mentioned that many detailed models of transmission networks have a meshed core with ``dangling'' appendages like IEEE 118-bus system.}.  For a clearer demonstration, we remove these dangling bridges and replace them by power injections at the corresponding endpoints in the giant block that equal
the power flows on these bridges. 
In addition, we switch off three transmission lines to create more non-bridge blocks to better illustrate the block diagonal structure of the LODF matrix. The resulting network is composed of 4 non-bridge blocks with 4, 13, 45 and 110 edges, connected by one bridge block and two cut vertices.

In \cite{LiangGuo2020,zocca2020spectral} we explore ways to judiciously switch off a small number of transmission lines to create more blocks for localizing failures. 
In this paper, however, we focus on characterizing the intrinsic properties of a power network itself rather than the design for network reconfiguration and dynamic controller. 

\subsection{LODF Matrix}
In our experiments, the system parameters are taken from the Matpower Package \cite{zimmerman2011matpower} and we calculate the LODF matrix $K=(K_{l\hat l}, l\in \calE, \hat l\in \calE)$ that describes the impact on other lines of a single line outage. For non-bridge outages, we directly calculate the LODF as defined in equation~(4) in Part I of this paper. For bridge outages that create islands, we adopt the proportional control and assume all nodes are participating with $\alpha_k = \alpha$ for all $k$.

We visualize the magnitude of LODF matrix in Fig.~\ref{fig:LODF-118}(a) by means of a heatmap, in which we reorder the lines based on the block they belong to. Specifically, we stack bridge blocks first, followed by non-bridge blocks in nondecreasing order of the block size. In addition, we set a color limit for better visualization so that $\abs{K_{l\hat l}} \leq 0.005$ maps to dark blue and $\abs{K_{l\hat l}} \geq 1$ maps to yellow. In Fig.~\ref{fig:LODF-118}(b), we plot the same heatmap under the AC power flow equations (the  four  yellow  solid  line  represent  four  transmission lines  whose  failure  lead  to  non-convergent  AC  power  flow  equations), where the LODF is computed directly from its definition $K_{l \hat l}=\frac{\Delta f_l}{f_{\hat l}}$.

In both the DC and AC case, the global effect of a bridge failure is clearly visible in the first column (since only one line is a bridge) of the LODF matrices in Fig.3 (a) and (b). Since almost all the entries of that column are non-zero, almost all surviving lines will be impacted by the failure of that bridge. 
For non-bridge failures, the LODF matrix in the DC case exhibits a clear block diagonal structure. 
In the AC case, however, the cross-block entries are not necessary zero, but they are relatively small. Moreover, the LODF within a block are similar for both cases, indicating that the LODF computed from DC model can be a good approximation for AC model. We further remark that the LODFs within a block can be small, but they are strictly nonzero in all cases, confirming our result in Theorem 7 from Part I.

\subsection{Influence Graph}
We further visualize the transmission line failure propagation patterns using an \emph{influence graph}. Despite being similar in concept to that of \cite{hines2017cascading,hines2013dual},  unlike these influence graphs, which are based on a probabilistic failure models, our influence graph is simply a visualization of the LODFs $K_{l \hat l}$ that we superimpose on the original network topology.
The IEEE 118-bus network topology is depicted in blue in Fig.~\ref{fig:inf_graph}. The corresponding influence graph has these transmission lines (in blue) as nodes and connect any two transmission lines $l$ and $\hat l$ in the influence graph (with a grey edge) if the corresponding LODF satisfies $|K_{l\hat l}|\geq 0.005$. As Fig.~\ref{fig:inf_graph} shows, the impact a non-bridge outages are ``blocked" by  cut-vertices or  bridges. Specifically, there are no edges connecting transmission lines that belong to different blocks, as predicted by our theory.

\section{Conclusion}\label{section:conclusion}


In Part II of this work, we make use of the spectral representation of power redistribution developed in Part I to provide a characterization of line failure localizability when the initial failure disconnects the original network. This, together with our results in Part I, establishes a mathematical theory that covers all initial failure scenarios and reveals how a general power system responds to such disturbances. A case study on the IEEE 118-bus test network corroborates the block-diagonal structure predicted by our theory, even when the system is under full AC power flow equations.

This work can be built upon in several ways:
(a) Some practical power systems have been operating in partitioned mode. However, their tie-lines are high voltage DC (HVDC) transmission lines. One major feature of HVDC lines is that they are sensitive to voltage disturbances \cite{hvdc2013} so outages inside a block can cause HVDC outages, and further triggering outages in other blocks. More study is needed to fully understand line failure localizability for such systems. (b) The network block decomposition has potential application in power systems planning.
It can also be used as a corrective action immediately after a contingency, similar to
controlled islanding, to prevent failure propagation.
It would be interesting to understand the tradeoffs of post-contingency corrective block configuration and controlled islanding and how they can be synergistically integrated for failure mitigation. (c) Our model builds upon DC power flow dynamics, which are accurate for small deviations but less so under large disruptions. Yet, our preliminary simulations suggest a strong underlying structure that connects the gap
between DC and AC models in the context of line failures.  
It would be interesting to understand this structure and develop bounds on the distance 
between DC and AC predictions. 
(d) It is possible to integrate fast-timescale frequency control into our framework to provide a control strategy with provably optimal localization and mitigation properties. It would be 
useful to see what the best way to deploy such a strategy in practical systems would be (see \cite{LiangGuo2020,zocca2020spectral} for more details). 

\bibliographystyle{IEEEtran}
\bibliography{biblio,PowerRef-201202}

\iftoggle{isreport}{
\section{Appendix: proofs}
\label{section:proofs}
\subsection{Proof of Theorem \ref{ch:cf; subsec:bridgeoutage; theorem:simplepath}} \label{proof:balancing_bus}
The proof is similar to that for the Simple Loop Criterion in Part I. Recall from Part I that the PTDF $D_{l, k\hat j}$ at participating buses in 
\eqref{ch:cf; subsec:bridgeoutage; eq:Klhatl.1} is given by:
$$
\frac{B_{l}} {\sum_{H\in \calT_\calE}\, \beta(H)} \left( \sum_{H\in \calT(ik, j\hat j)}\, \beta(H) \ - \ \sum_{H\in \calT(i\hat j, jk)}\, \beta(H) \right)
$$
Note that the spanning forests $H$ in $\calT_\calE$, $\calT(ik, j\hat j)$ and $T(i\hat j, jk)$ are spanning forests of the
pre-contingency network, not just the post-contingency island.

Suppose $D_{l, k\hat j}\neq 0$.  Then either 
$\calT(ik, j\hat j)$ or $\calT(i\hat j, jk)$ or both is nonempty.
Suppose without loss of generality that
 there is a spanning forest $H\in \calT(ik, j\hat j)$.  Then there is a path in $H$ from buses $\hat j$ to $j$ and another
vertex-disjoint path from buses $i$ to $k$.  Joining these two paths with the line $l = (i, j)$ creates a simple path from $\hat j$ 
to $k$ that contains line $l$, as desired.

Conversely suppose there is a simple path in $\calG$ from $\hat j$ to a participating bus $k'\in \calN$ that
contains $l = (i,j)$.  We will show that $g(B) := \sum_{k} D_{l, k\hat j}(B)\alpha_k$ in 
\eqref{ch:cf; subsec:bridgeoutage; eq:Klhatl.1} as a polynomial in
the susceptances $B$ of \emph{all} lines in the pre-contingency network
 is not identically zero, and hence $\mu(g(B+\omega) = 0)= 0$.
Since the pre-contingency network is connected the simple path in the island $\calG$ from buses $\hat j$ to $k'$ that contains $l = (i,j)$ can be
extended into a spanning tree of the pre-contingency network.  Suppose without loss of generality that, on this spanning tree, the path 
from $\hat j$ to $i$ contains $j$, i.e., the path from $k'$ to $\hat j$ is of the form 
$k' \rightsquigarrow i \rightarrow j \rightsquigarrow \hat j$.  Then removing the edge $l$ from this spanning tree 
creates a spanning forest $H$ in $\calT(ik', j\hat j)$ that contains exactly two trees, denoted by $H_{ik'}$ connecting 
buses $i, k'$ and $H_{j \hat j}$ connecting buses $j, \hat j$.
Following a similar argument in our proof for the Simple Cycle Criterion, for $H\not\in \calT(i\hat j, jk')$ the term $\beta(H)$ is not canceled by
a negative term in $D_{l, k' \hat j}$.
Moreover we claim that $H\not\in \calT(i\hat j, jk)$ for all other participating buses $k$ and therefore $\beta(H)$ is 
not canceled by negative terms from other $D_{l, k \hat j}$ in $g(B)$ either.
To see this, if $H\in \calT(i\hat j, jk)$ for a participating bus $k$ then there must be a path in $H$ connecting buses $i$ to
$\hat j$, but this path then connects tree $H_{ik'}$ to tree $H_{j\hat j}$ into a spanning tree.  This contradicts that the spanning forest
$H \in \calT(ik', j\hat j)$ consists of two distinct trees.
Hence $g(B)$ is not a zero polynomial and $K_{l\hat l}\neq 0$ $\mu$-almost surely.
\qed

\subsection{Proof of Lemma \ref{lemma:participating_region}}\label{proof:participating_region}
Let $\ol{\calC}$ be the block that contains $l$. Since $\calE_k$ is a participating region, we know there exists a bus within $\calC$, say $n_1$, that participates the power balance and is not a cut vertex. Recall that any path from $j(\hat{l})$ to $\calC$ must go through a common cut vertex in $\calC$ \cite{harary1969graph}, say $n_e$. Now by adding an edge between $n_e$ and $n_1$ (if such edge did not originally exist), the resulting block $\calC'$ is still 2-connected. Thus there exists a simple cycle in $\calC'$ that contains the edge $(n_e,n_1)$ and $l:=(i,j)$, which implies we can find two disjoint paths $P_1$ and $P_2$ connecting the endpoints of these two edges. Without loss of generality, assume $P_1$ connects $n_e$ to $i$ and $P_2$ connects $n_1$ to $j$. By concatenating the path from $j(\hat{l})$ to $n_e$, we can extend $P_1$ to a path $\tilde{P}_1$ from $j(\hat{l})$ to $i$, which is still disjoint from $P_2$. Now, by adjoining $\hat{e}$ to $\tilde{P}_1$ and $P_2$, we can construct a path from $j(\hat{l})$ to $n_1$ that passes through $l$. The Simple Path Criterion then implies $\mu(\Delta f_{\hat{e}}\neq 0)=1$. \qed

\subsection{Proof of Theorem \ref{ch:cf; sec:localization; subsec:cutset; thm:cutset}}
The first part of Theorem \ref{ch:cf; sec:localization; subsec:cutset; thm:cutset} is proved by analyzing the post-contingency network $(\calN, \calE \setminus F)$ using the matrix $A_{-F}$, while the second part leverages the relation between the matrices $A$ for the pre- and post-contingency networks.
Similar to our analysis of GLODF, taking the difference between \eqref{ch:cf; subsec:bridgeoutage; eq:DCPFpost.2}
and \eqref{ch:cf; subsec:bridgeoutage; eq:DCPFpre.2}  shows that the branch flow changes
$\Delta f_{-F} := \tilde f_{-F} - f_{-F}$ satisfy the post-contingency DC power flow equations:
\begin{IEEEeqnarray*}{rCl}
    C_{F} f_{F}  + \left( \Delta p_\alpha - \Delta p_\text{tie} \right) & = & C_{-F}  \Delta f_{-F},\\
    \Delta f_{-F} & = & B_{-F} C_{-F}^T \left( \tilde \theta - \theta \right).
\end{IEEEeqnarray*}
Recall that $A_{-F} $ is defined in terms of the inverse of the reduced Laplacian matrix $L_{-F} = C_{-F} B_{-F} \, C_{-F}^T$
for the post-contingency network. The branch flow changes are given by
\begin{IEEEeqnarray*}{rCl}
    \Delta f_{-F} &  = & B_{-F} C_{-F}^T A_{-F}\left( C_{F} f_{F} + \Delta p_\alpha - \Delta p_\text{tie} \right) \\
    & = & \underbrace{ B_{-F}  C_{-F}^T A_{-F} C_{F} }_{K^{F}}  f_{F} \\
    && + \underbrace{ B_{-F} \, C_{-F}^T \, A_{-F} }_{\hat D^F} \left( \Delta p_\alpha - \Delta p_\text{tie} \right),
\end{IEEEeqnarray*}
where the GLODF $K^{F} = B_{-F} \, C_{-F}^T \, A_{-F}\, C_{F}$ is defined by Theorem 8 in Part I
and $\hat D^F$ is defined in \eqref{ch:cf; sec:localization; subsec:cutset; eq:DF.1a}.
From \eqref{ch:cf; subsec:bridgeoutage; eq:Galpha.1b} and \eqref{ch:cf; subsec:bridgeoutage; eq:P} it follows that
\begin{eqnarray}
    \Delta p_\alpha - \Delta p_\text{tie} = 
    \sum_{\hat l\in F_\text{tie}} f_{\hat l} \sum_{k\in \calN} \alpha_k \left( e_k - e_{j(\hat l)} \right).
\label{ch:cf; subsec:bridgeoutage; eq:Deltapa-pt}
\end{eqnarray}
Hence the branch flow changes are the sum of the impacts of internal line outages in $F$ and 
tie line outages in $F_{\text{tie}}$:
\begin{equation*}
    \Delta f_{-F} = \underbrace{ K^{F} \, f_{F} }_{ \text{int. line $F$ outage} }
	 + 
	\underbrace{	
	\sum_{\hat l\in F_\text{tie}} f_{\hat l} \sum_{k\in \calN} \alpha_k 
	\hat D^F \left( e_k - e_{j(\hat l)} \right)
	}_{ \text{tie line $F_\text{tie}$ outage} }
\label{ch:cf; subsec:bridgeoutage; eq:DeltaP.1}
\end{equation*}
This proves identity~\eqref{ch:cf; sec:localization; subsec:cutset; eq:cutset.1} of Theorem \ref{ch:cf; sec:localization; subsec:cutset; thm:cutset}. 

We now show that \eqref{ch:cf; sec:localization; subsec:cutset; eq:DF.1b} is an alternative representation of \eqref{ch:cf; sec:localization; subsec:cutset; eq:DF.1a}. To do so, we express the matrix $\hat D^F$ in terms of the matrices of the pre-contingency network. In particular, we relate $A_{-F}$ with $A$ using matrix inversion lemma as follows:
\begin{IEEEeqnarray}{rCl}
    \hat D^F & = & B_{-F} C_{-F}^T A_{-F} \nonumber \\
    & = & B_{-F} C_{-F}^T A  \nonumber\\
    && + B_{-F} C_{-F}^T AC_F \left(I - B_FC_F^TAC_F\right)^{-1}B_F C_F A \nonumber \\
    & = & B_{-F} C_{-F}^T A + K^F B_F C_F A. \nonumber \hfill \qedhere
\end{IEEEeqnarray}

\qed




}{}

\begin{IEEEbiography}[{\includegraphics[width=1in,height=1.25in,clip,keepaspectratio]{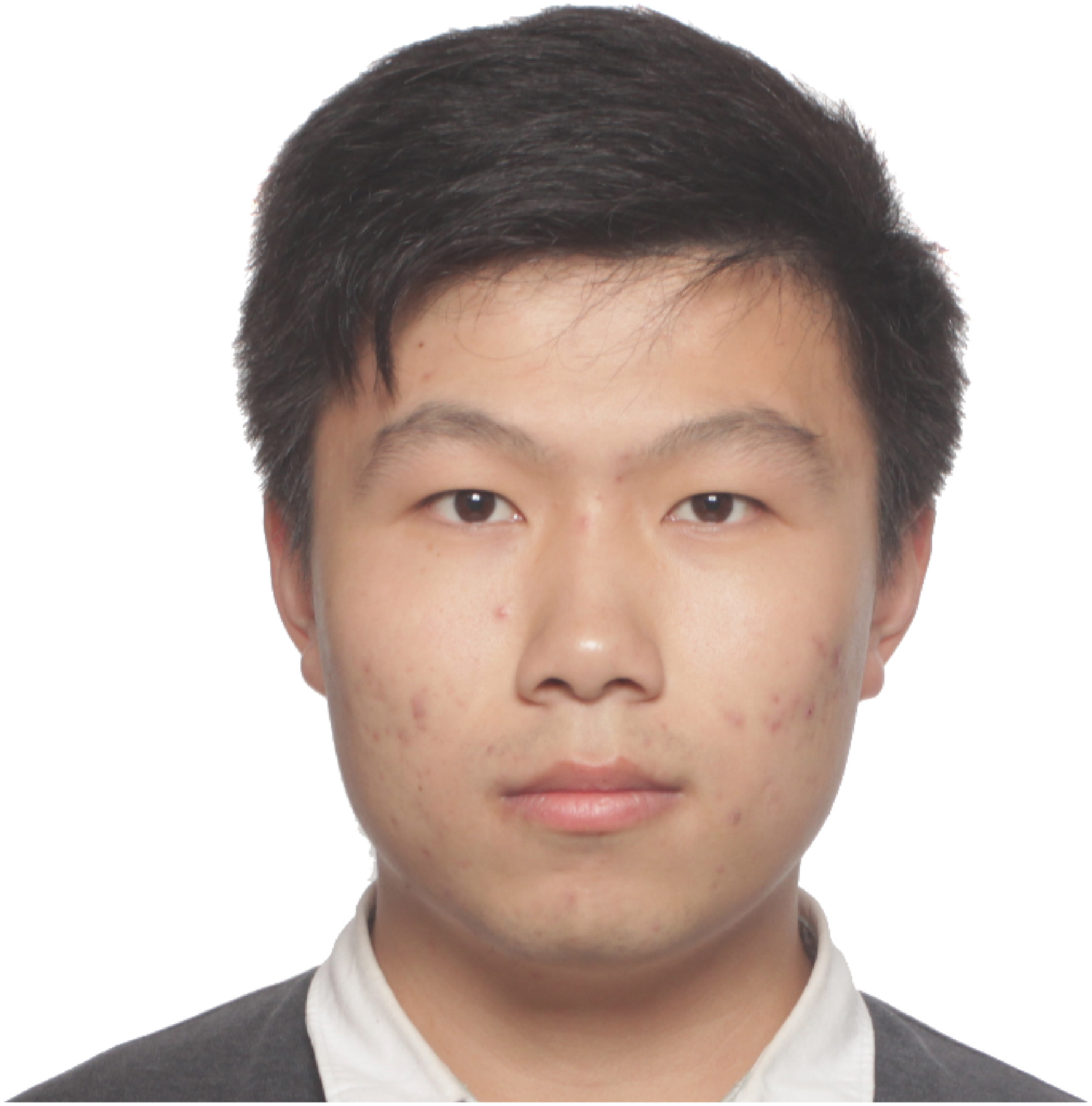}}]{Linqi Guo}
 received his B.Sc.~in Mathematics and B.Eng.~in Information Engineering from The Chinese University of Hong Kong in 2014, and his Ph.D.~in Computing and Mathematical Sciences from California Institute of Technology in 2019. His research is on the control and optimization of networked systems, with focus on power system frequency regulation, cyber-physical network design, distributed load-side control, synthetic state estimation and cascading failure analysis.
\end{IEEEbiography}

\begin{IEEEbiography}[{\includegraphics[width=1in,height=1.25in,clip,keepaspectratio]{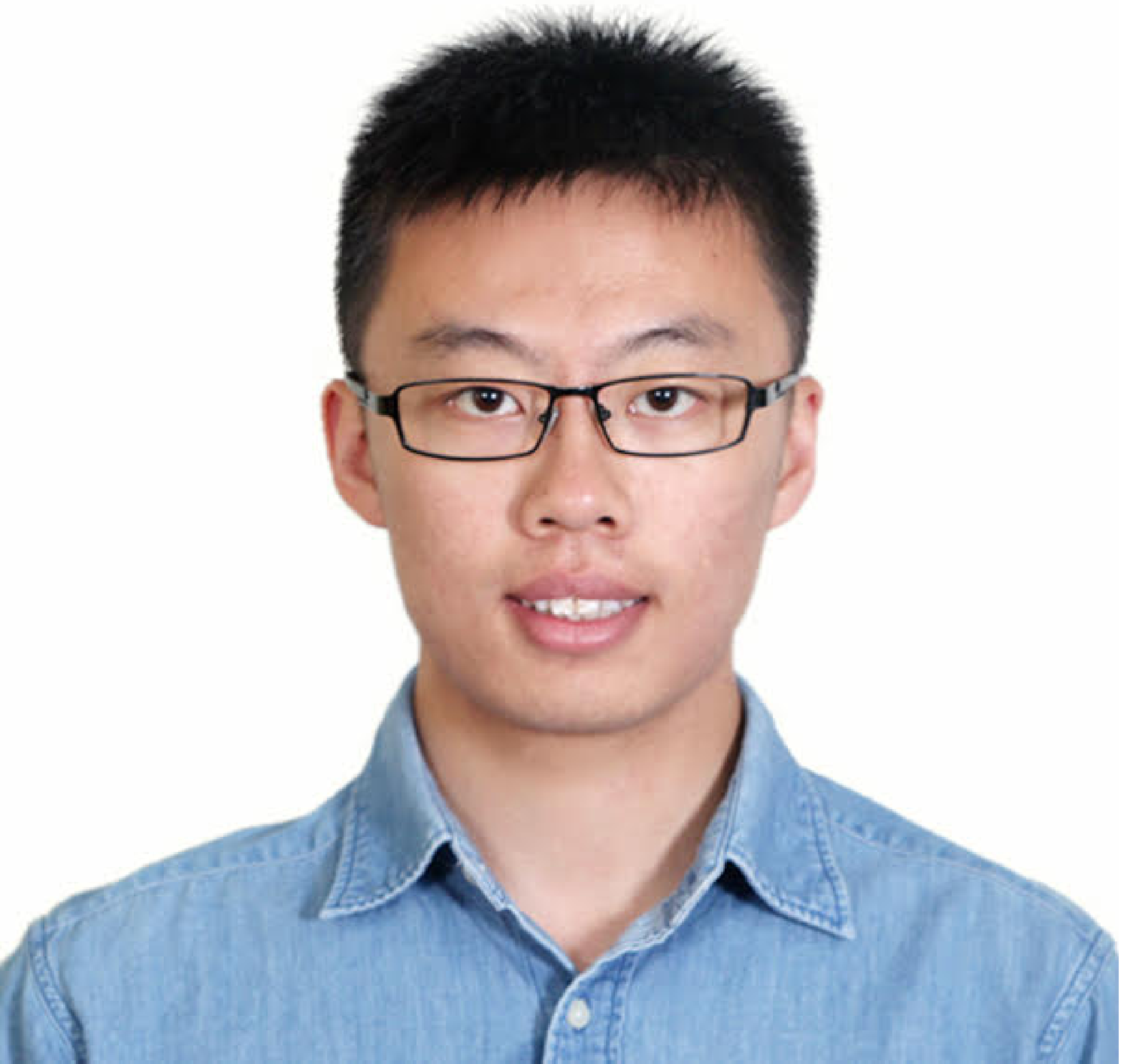}}]{Chen Liang (S’19)}
 received the B.E. degree in automation from Tsinghua University, Beijing, China, in 2016. He is currently pursuing the Ph.D. degree in Computing and Mathematical Sciences with the California Institute of Technology, Pasadena, CA, USA. 
His research interests include graph theory, mathematical optimization, control theory, and their applications in cascading failures of power systems.
\end{IEEEbiography}

\begin{IEEEbiography}[{\includegraphics[width=1in,height=1.25in,clip,keepaspectratio]{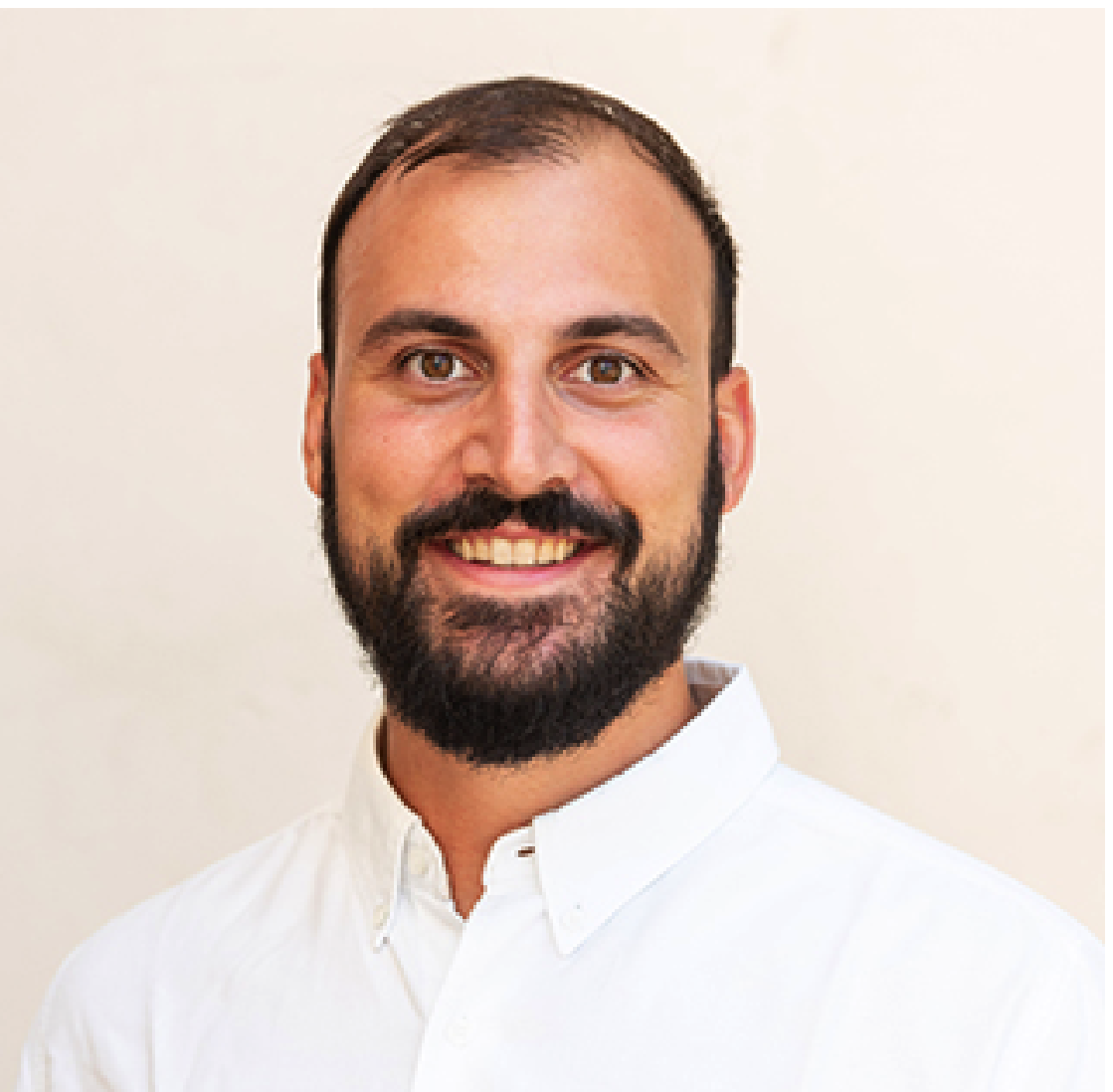}}]{Alessandro Zocca}
 received his B.Sc. in mathematics from the University of Padua, Italy, in 2010, his M.A.St. in mathematics from the University of Cambridge, UK, in 2011, and his Ph.D. degree in mathematics from the University of Eindhoven, The Netherlands, in 2015. He then worked as postdoctoral researcher first at CWI Amsterdam (2016-2017) and then at California Institute of Technology (2017-2019), where he was supported by his personal NWO Rubicon grant. Since October 2019, he has a tenure-track assistant professor position in the Department of Mathematics at the Vrije Universiteit Amsterdam. His work lies mostly in the area of applied probability and optimization, but has deep ramifications in areas as diverse as operations research, graph theory, algorithm design, statistical physics, and control theory. His research focuses on dynamics and rare events on large-scale networked systems affected by uncertainty, drawing motivation from applications to power systems and wireless networks.
\end{IEEEbiography}

\begin{IEEEbiography}[{\includegraphics[width=1in,height=1.25in,clip,keepaspectratio]{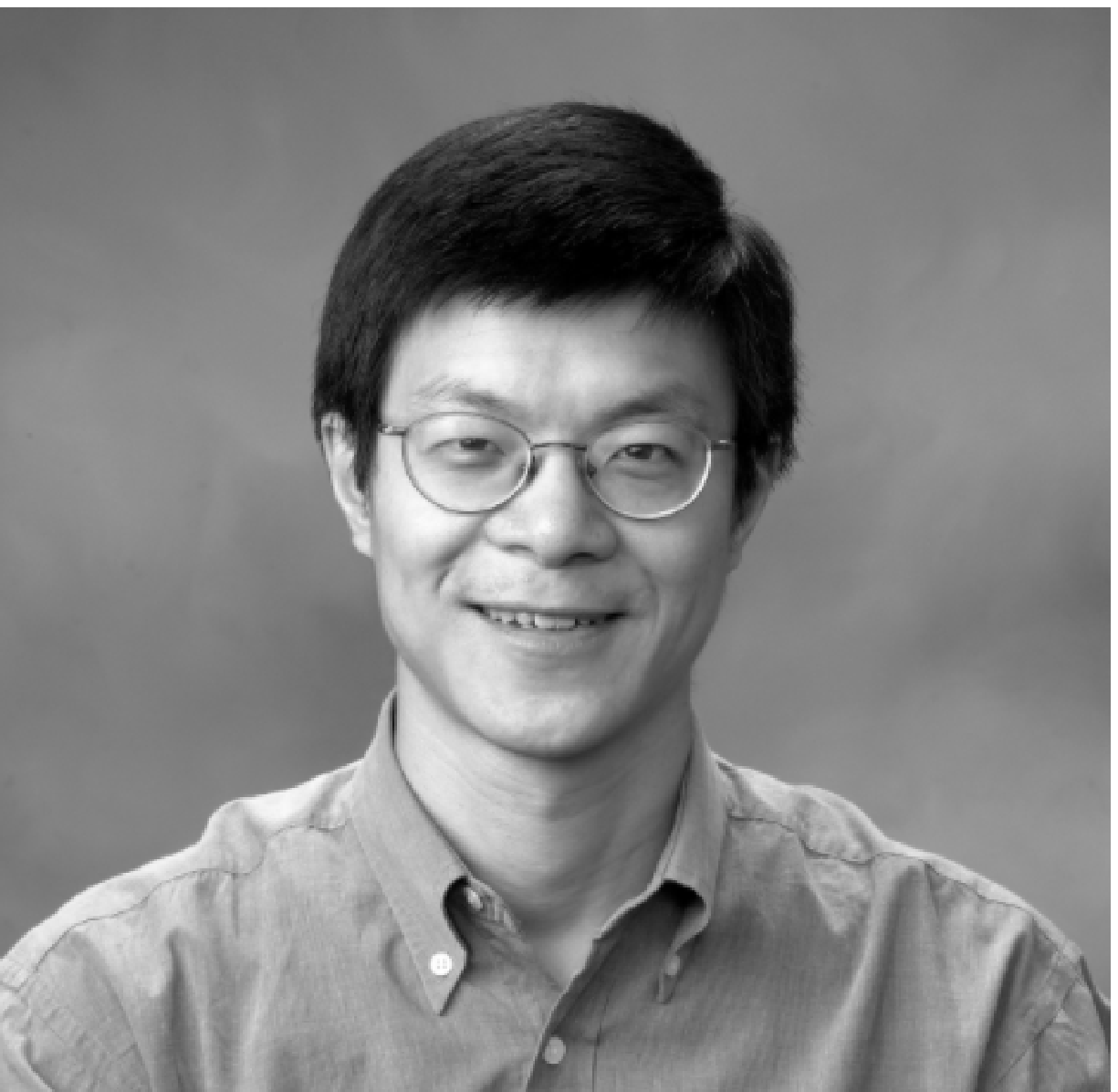}}]{Steven H. Low (F’08)}
is the F.~J.~Gilloon Professor of the Department of Computing \& Mathematical Sciences and the Department of Electrical Engineering at Caltech. Before that, he was with AT\&T Bell Laboratories, Murray Hill, NJ, and the University of Melbourne, Australia. He has
held honorary/chaired professorship in Australia, China and Taiwan. He was a co-recipient of IEEE best paper awards and is a Fellow of both IEEE and ACM. He was known for pioneering a mathematical theory of Internet congestion control and semidefinite relaxations of optimal power flow problems in smart grid.  He received his B.S. from Cornell and PhD from Berkeley, both in electrical engineering.
\end{IEEEbiography}

\begin{IEEEbiography}[{\includegraphics[width=1in,height=1.25in,clip,keepaspectratio]{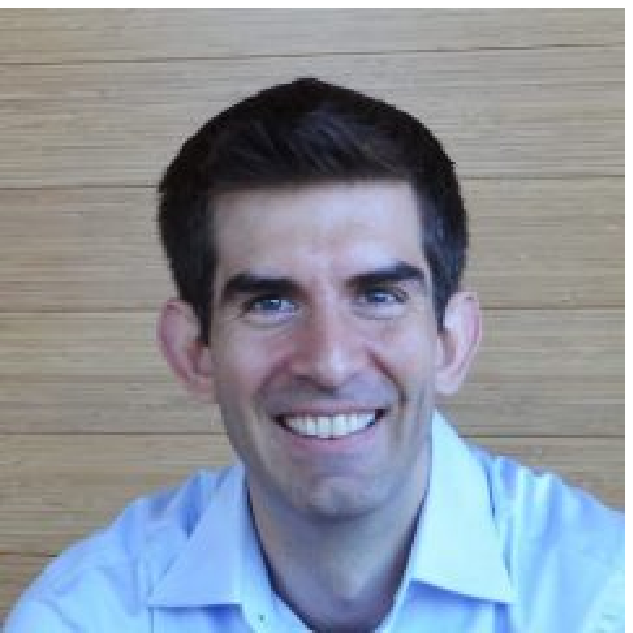}}]{Adam Wierman}
is a Professor in the Department of Computing and Mathematical Sciences at the California Institute of Technology. He received his Ph.D., M.Sc.~and B.Sc.~in Computer Science from Carnegie Mellon University in 2007, 2004, and 2001, respectively, and has been a faculty at Caltech since 2007. He is a recipient of multiple awards, including the ACM SIGMETRICS Rising Star award, the IEEE Communications Society William R. Bennett Prize, and multiple teaching awards. He is a co-author of papers that have received best paper awards at a wide variety of conferences across computer science, power engineering, and operations research including ACM Sigmetrics, IEEE INFOCOM, IFIP Performance, and IEEE PES.
\end{IEEEbiography}

\end{document}